\documentstyle[12pt]{article}

\makeatletter
\def\eqnarray{\stepcounter{equation}\let\@currentlabel=\theequation
\global\@eqnswtrue
\global\@eqcnt\z@\tabskip\@centering\let\\=\@eqncr
$$\halign to \displaywidth\bgroup\@eqnsel\hskip\@centering
  $\displaystyle\tabskip\z@{##}$&\global\@eqcnt\@ne
  \hfil$\displaystyle{\hbox{}##\hbox{}}$\hfil
  &\global\@eqcnt\tw@ $\displaystyle\tabskip\z@
  {##}$\hfil\tabskip\@centering&\llap{##}\tabskip\z@\cr}
\@addtoreset{equation}{section}
  \def\theequation{\thesection.\arabic{equation}}
\makeatother

\advance \textheight by \topskip
\begin{document}

\title{$R$-deformed Heisenberg algebra, anyons and $d=2+1$
supersymmetry}
\author{Mikhail S. Plyushchay
{}\thanks{e-mail: plyushchay@mx.ihep.su}\\
\smallskip\\
{\it Departamento de Fisica}\\
{\it Universidade Federal de Juiz de Fora}\\
{\it 36036-330 Juiz de Fora, MG Brazil}\\
{\it and}\\
{\it Institute for High Energy Physics}\\
{\it Protvino, Moscow Region, 142284 Russia}}

\date{}

\maketitle

\begin{abstract}
A universal minimal spinor set of linear differential equations
describing anyons and ordinary integer and half-integer spin
fields is constructed with the help of deformed Heisenberg
algebra with reflection.  The construction is generalized to
some $d=2+1$ supersymmetric field systems. Quadratic and linear
forms of action functionals are found for the universal minimal
as well as for supersymmetric spinor sets of equations.  A
possibility of constructing a universal classical mechanical
model for $d=2+1$ spin systems is discussed.

\centerline
{\bf Mod. Phys. Lett. A12 (1997) 1153-1164}
\end{abstract}

\section{Introduction}
There are two different field-theoretical approaches to the
description of anyons
\cite{a1,a2,a3} as fundamental particle-like objects. 
One of them uses Chern-Simons gauge field whose role is to
change spin and statistics of the charged matter field minimally
coupled to it \cite{a3,cs1,cs2}.  The initial formulation of the
Chern-Simons construction has a local character, but the final
gauge-invariant composite field carrying fractional spin and
obeying fractional statistics turns out to be of an essentially
nonlocal nature
\cite{cs1,cs2}.  Another is the so called group-theoretical approach
\cite{gr1}--\cite{chou}, 
which aims at describing anyons by means of the methods
applicable for ordinary integer and half-integer spin fields (in
what follows, spin-$j$ fields).  This approach can be formulated
starting either from the construction of the appropriate field
equations \cite{jn}--\cite{gr6}, 
or just from modelling fractional spin
particles (further on, fractons) at the level of classical
mechanics with a subsequent quantization of the theory 
\cite{gr1,sv,gr5}.
However, it seems that spin-$j$ particles and fractons have
quite different nature within the latter approach too.  Indeed,
spin degrees of freedom of spin-$j$ particles are described by
anticommuting grassmann (odd) \cite{pseud} or paragrassmann
\cite{para}
variables in pseudoclassical mechanics, whereas fractons are
described by constrained dynamics on extended ``plain" phase
space with commuting (even) spin variables \cite{gr1,sv,gr5}.
Alternatively, fractional spin particles require using a
monopole-like symplectic structure  on the minimal ``curved"  phase
space
containing no spin variables at all \cite{jn,gr5,chou}.  At the
level of field equations one uses {\it finite-dimensional
non-unitary} representations of $\overline{SO(2,1)}$ for
describing spin-$j$ fields, whereas {\it infinite-dimensional
unitary} representations of the universal cover of $d=2+1$
Lorentz group have to be employed in the case of
anyons\footnote{Using the term ``anyons", it is necessary to
have in mind that the spin-statistics relation has not yet been
proved for fractional spin particles within a group-theoretical
approach.} \cite{jn,gr2,gr3}.

In this letter we shall reveal a general algebraic structure
unifying the cases of spin-$j$ fields and fractons.  Namely, we
shall construct a universal covariant set of linear differential
equations describing either spin-$j$ fields or fractional spin
fields.  This underlying algebraic structure will be
given by the deformed Heisenberg algebra involving the
reflection operator $R$ \cite{ok}.  
Here we shall use the universality of
the $R$-deformed Heisenberg algebra (RDHA) which has
been established recently \cite{uni}: 
this algebra is related simultaneously to
(generalized) parabosons and to (deformed) parafermions
supplying us with irreducible unitary infinite-dimensional and
non-unitary
finite-dimensional representations of $osp(1|2)$ superalgebra.
The set of equations which will be constructed is a minimal spinor
set
of equations ${\cal Q}_\alpha\Psi=0$ with 
linear (in $\partial_\mu$) differential spinor operator ${\cal
Q}_\alpha$ to be realized via the generators of RDHA. 
The hidden $osp(1|2)$ superalgebraic structure of RDHA will also
allow us to construct spinor linear sets of equations
describing some (2+1)-dimensional supersymmetric systems.

The paper is organized as follows.  In section 2 we formulate the
problem of constructing a minimal universal covariant set of
linear differential equations and describe a universal
$osp(1|2)$ superalgebraic hidden structure of RDHA necessary for
realizing the construction.  In section 3 we solve the
formulated problem.  Section 4 is devoted to generalization for
the case of supersymmetric spinor sets of linear differential
equations.  Quadratic and linear variants of field actions
giving rise to corresponding universal or SUSY sets of equations
are presented in section 5.  Section 6 is devoted to concluding
remarks.  A short Appendix contains some relations necessary for the
constructions realized in the main text of the paper.

\section{R-deformed Heisenberg algebra and $osp(1|2)$} 

In 2+1 dimensions, the fields $\Psi=\Psi^n(x)$ obeying two
equations
\begin{equation}
(P^2+m^2)\Psi=0,\quad
(PJ-sm)\Psi=0
\label{a1}
\end{equation}
describe one-particle states realizing irreducible
representations of $d=2+1$ quantum mechanical 
Poincar\'e group \cite{a2,jn,gr3}.  Here
$P_\mu=-i\partial_\mu$, $\mu=0,1,2,$ $J_\mu$ is
translation-invariant part of the total angular momentum vector
operator
$
M_\mu=-\epsilon_{\mu\nu\lambda}x^\nu P^\lambda + J_\mu,
$
$
[J_\mu,J_\nu]=-i\epsilon_{\mu\nu\lambda}J^\lambda,
$
and we use the metric $\eta_{\mu\nu}=diag(-,+,+)$ and
the totally antisymmetric tensor $\epsilon_{\mu\nu\lambda}$,
$\epsilon^{012}=1$.  Quadratic Klein-Gordon equation is the
consequence of the second linear spin equation only under the
choice of two-dimensional spinor ($n=1,2$, $s=\pm 1/2$) or
three-dimensional vector ($n=1,2,3$, $s=\pm 1$) representations.
For any other finite-dimensional or any infinite-dimensional
representation of $so(2,1)$, these equations fix independently
two Casimir operators of $d=2+1$ Poincar\'e group \cite{jn,gr3}.
Due to their independence, equations (\ref{a1}) are not very
suitable as a starting point for the construction of
corresponding field action and subsequent quantization 
of fractons and spin-$j$ fields with $j>1$ \cite{jn,gr3,gr6}.

Here we shall use the Dirac's idea to generate
Klein-Gordon equation as an integrability condition
of some covariant set of linear differential equations
\cite{dir}.
Recently, such an idea was employed 
to describe fractons as well as $d=2+1$ spin-$j$ fields 
\cite{sv,gr3,gr4,gr6}.
However, the previous variants of linear differential 
equations describe either only fractons \cite{sv,gr4,gr6}, or 
both $d=2+1$ spin-$j$ fields and anyons, but
in a non-minimal way \cite{gr3}.
A spinor set of two linear differential
equations to be constructed here will generate both equations
(\ref{a1}) for the case of any representation of $so(2,1)$. As
a consequence, it will give us a universal minimal covariant
description of fractons and spin-$j$ fields.  Since the
construction will be based on $osp(1|2)$ superalgebra realized
universally in terms of the $R$-deformed Heisenberg algebra
generators, for the sake of completeness of exposition
we describe briefly representations of the latter
algebra and related realization of $osp(1|2)$ generators.

The deformed Heisenberg algebra with reflection \cite{ok,uni}
is given by the
generators $a^-$, $a^+$, $R$, and $1$, which satisfy the
(anti)commutation relations
\begin{equation}
[a^-,a^+]=1+\nu R,\quad
R^2=1,\quad 
\{a^\pm,R\}=0,
\quad
[1,a^\pm]=[1,R]=0,
\label{a3}
\end{equation}
where $\nu\in {\bf R}$ is a deformation parameter and $R$ is 
a reflection operator.
In the case 
$
\nu>-1
$
albegra (\ref{a3}) has infinite-dimensional unitary
representations which can be realized on the Fock space with
complete orthonormal basis  of states
$|n\rangle=C_n(a^+)^n|0\rangle$, $n=0,1,\ldots,$
$a^-|0\rangle=0$, $\langle 0|0\rangle=1$,
$R|0\rangle=|0\rangle$, where
$C_n=([n]_\nu!)^{-1/2}$, $[n]_\nu !=\prod_{l=1}^n [l]_\nu$,
$[l]_\nu=l+\frac{1}{2}(1-(-1)^l)\nu$.  The reflection operator
acts as $R|n\rangle=(-1)^n|n\rangle$, introducing $Z_2$-grading
structure in the space of states, $
R|k\rangle_\pm=\pm|k\rangle_\pm, $ $ |k\rangle_+=|2k\rangle, $ $
|k\rangle_-=|2k+1\rangle, $ $ k=0,1,\ldots.  $ The even,
`$+$', and odd, `$-$', subspaces can be singled out by the
projectors $
\Pi_\pm=\frac{1}{2}(1\pm R),
$
$
\Pi_\pm^2=\Pi_\pm,
$
$
\Pi_+\Pi_-=0,
$
$
\Pi_++\Pi_-=1.
$

Algebra (\ref{a3}) has also $(2r+1)$-dimensional 
representations for the values of deformation parameter 
$
\nu=-(2r+1),\quad
r=1,2,\ldots
$
\cite{uni}.
These representations are characterized by the relations
$
a^{-(2r+1)}=a^{+(2r+1)}=0
$
being specific for paragrassmann algebras \cite{fiks}.
They  can be realized as matrix representations
with diagonal operator $R=diag(+1,-1,+1,\ldots,
-1,+1)$, and with operators $a^\pm$ realized as
$
(a^+)_{ij}=A_j\delta_{i-1,j},\quad
(a^-)_{ij}=B_i\delta_{i+1,j},
$
where
$
A_{2k+1}=-B_{2k+1}=\sqrt{2(r-k)},
$
$
k=0,1,\ldots,r-1,
$
$
A_{2k}=B_{2k}=\sqrt{2k},
$
$
k=1,\ldots,r.
$
Operators $a^+$ and $a^-$ are conjugate,
$(\Psi_1,a^-\Psi_2)^*=(\Psi_2,a^+\Psi_1)$, 
with respect to
the indefinite scalar product given by
\begin{equation}
(\Psi_1,\Psi_2)=\bar{\Psi}_{1n}\Psi_{2}^n,
\quad
\bar{\Psi}_n=\Psi^{*k}\hat{\eta}_{kn},
\label{a4}
\end{equation}
where $\Psi^n=\langle n|\Psi\rangle$ and
$\hat{\eta}=diag(1, -1, -1, +1, +1,\ldots,
(-1)^{r-1}, (-1)^{r-1}, (-1)^r, (-1)^r)$
is indefinite metric operator.
Linear combinations
$
{\cal L}_1=\frac{1}{\sqrt{2}}(a^++a^-),$
${\cal L}_2=\frac{i}{\sqrt{2}}(a^+-a^-)$,
being the $R$-deformed coordinate and momentum
operators,
$[{\cal L}_\alpha,{\cal L}_\beta]=i\epsilon_{\alpha\beta}
(1+\nu R)$, 
$\epsilon_{\alpha\beta}=-\epsilon_{\beta\alpha},$
$\epsilon_{12}=1$,
are hermitian in the case of infinite-dimensional representations
and self-conjugate with respect to scalar product (\ref{a4})
in the case of finite dimensional representations of RDHA.

Operators ${\cal L}_\alpha$
together with quadratic operators $J_\mu$,
$
J_0\equiv\frac{1}{4}\{a^+,a^-\},
$
$
J_1\pm i J_2=J_\pm=\frac{1}{2}(a^\pm)^2,
$
form the set of
generators of $osp(1|2)$ superalgebra:
$
\{{\cal L}_\alpha,{\cal L}_\beta\}
=4i(J\gamma)_{\alpha\beta},
$
$
[J_\mu,J_\nu]=-i\epsilon_{\mu\nu\lambda}J^\lambda$,
$
[J_\mu,{\cal L}_\alpha]=
\frac{1}{2}(\gamma_\mu)_\alpha{}^{\beta}{\cal L}_\beta.
$
These (anti)commutation relations mean that
$J_\mu$ are even and ${\cal L}_\alpha$ are odd generators 
of the superalgebra and that the components of ${\cal L}_\alpha$
form $so(2,1)$ spinor.
Here $\gamma$-matrices, satisfying 
the relations 
$\gamma^{\mu}\gamma^{\nu}=-\eta^{\mu\nu}+i\epsilon^{\mu\nu\lambda}
\gamma_{\lambda}$,
appear in the Majorana representation,
$
(\gamma^{0})_{\alpha}{}^{\beta}=-(\sigma^{2})_{\alpha}{}^{\beta},
$
$
(\gamma^{1})_{\alpha}{}^{\beta}=i(\sigma^{1})_{\alpha}{}^{\beta},
$
$
(\gamma^{2})_{\alpha}{}^{\beta}=i(\sigma^{3})_{\alpha}{}^{\beta},
$
and so, they satisfy also the relations 
$
\gamma^{\mu\dagger}_{\alpha\beta}=
-\gamma^{\mu}_{\alpha\beta},$
$\gamma^{\mu}_{\alpha\beta}=\gamma^{\mu}_{\beta\alpha}$.
Raising and lowering spinor 
indices is realized  by 
$\epsilon_{\alpha\beta}$ and by
the antisymmetric tensor $\epsilon^{\alpha\beta}$,
$\epsilon^{12}=1$: 
$f_{\alpha}=f^{\beta}\epsilon_{\beta\alpha},$ 
$f^{\alpha}=\epsilon^{\alpha\beta}f_{\beta}.$ 
The $osp(1|2)$ Casimir operator 
$
{\cal C}\equiv J^\mu J_\mu-\frac{i}{8}{\cal L}^\alpha {\cal L}_\alpha
$
takes the fixed value ${\cal C}=\frac{1}{16}(1-\nu^2)$ and
therefore,  every infinite- or finite-dimensional representation
of RDHA supplies us with corresponding irreducible
representation of $osp(1|2)$ superalgebra.

On the other hand, every such  representation is reducible
with respect to the action of the $so(2,1)\sim
sl(2,R)$ generators $J_\mu$:
$
J^2=J^\mu J_\mu=-\hat{\alpha}(\hat{\alpha}-1),
$
where
$
\hat{\alpha}=\frac{1}{4}(1+\nu R).
$
Therefore, $J_\mu$ act irreducibly on even, `$+$', and odd,
`$-$',  subspaces 
spanned by the states $|k\rangle_+$ and $|k\rangle_-$,
$
J^2|k>_\pm=-\alpha_\pm(\alpha_\pm-1)|k>_\pm,
$
where 
$\alpha_+=\frac{1}{4}(1+\nu)$,
$\alpha_-=\alpha_++1/2$
and 
$
J_0|k>_\pm=(\alpha_\pm+k)|k>_\pm,\quad k=0,1,\ldots.
$
In the case of infinite-dimensional representations of RDHA
($\nu>-1$), this  gives the direct sum of infinite-dimensional
unitary irreducible representations of $sl(2,R)$, ${\cal
D}^+_{\alpha_+}\oplus{\cal D}^+_{\alpha_-}$, being
representations of the so called discrete series with parameters
$\alpha_+>0$ and $\alpha_->1/2$ \cite{gr6}.
In the case of finite-dimensional representations 
of RDHA, we have the relations
$
J^2|l>_\pm=-j_\pm(j_\pm+1)|l>_\pm,
$
$j_+=r/2$, $j_-=(r-1)/2$,
where $l=0,1,\ldots,r$ for
$|l>_+$ 
and $l=0,1\ldots,r-1$ for
$|l>_-$.
Therefore, we have the direct sum of
spin-$j_+$ and spin-$j_-$ finite-dimensional representations
with $so(2,1)$ spin parameter shifted in $1/2$,
where the operator $J_0$ has the spectra
$j_0=(-j_+,-j_++1,\ldots,j_+)$ and $j_0=(-j_-,-j_-+1,\ldots,
j_-)$, respectively \cite{uni}.

As we shall see, the presence of two irreducible representations
of $so(2,1)$ with corresponding spin parameter shifted in $1/2$
will allow us to realize (2+1)-dimensional SUSY.

\section{Universal spinor set of equations}
Let us supplement $d=2+1$ Lorentz spinor operator
${\cal L}_\alpha$ with the set of spinor operators
$
{\cal P}_\alpha\equiv
(P\gamma)_\alpha{}^\beta {\cal L}_\beta,
$
$
{\cal J}_\alpha\equiv 
i{\cal L}_\beta\epsilon_{\mu\nu\lambda}P^\mu J^\nu
(\gamma^\lambda)_\alpha{}^{\beta}
$
and
$
{\cal L}_\alpha(PJ)
$
linearly depending on $P_\mu$.
Due to the identity relation 
$
{\cal L}_\alpha(PJ)+ \frac{1}{4}(3+\nu R)
{\cal P}_\alpha-{\cal J}_\alpha
\equiv 0,
$
only any two operators from the introduced 
$P_\mu$-dependent set can be
considered as independent spinor operators.  We shall use the
first two operators, ${\cal P}_\alpha$
and ${\cal J}_\alpha$, as independent ones.  
The introduced operators have the 
properties $ {\cal P}^\dagger_\alpha=-{\cal P}_\alpha, $ 
$ {\cal J}^\dagger_\alpha={\cal J}_\alpha-{\cal P}_\alpha, $ where we
mean the conjugation with respect to the positive definite
scalar product $\langle\Psi_1,\Psi_2\rangle=
\Psi^{*n}_{1}\Psi_{2}^n$ 
in the case $\nu>-1$,
and with respect to indefinite scalar product 
(\ref{a4}) for $\nu=-(2r+1)$, $r=1,2,\ldots$.

One considers the self-conjugate spinor operator
\begin{equation}
{\cal Q}_\alpha={\cal Q}^\dagger_\alpha=
R{\cal P}_\alpha +\epsilon m{\cal L}_\alpha,
\quad
\epsilon=\pm 1,
\label{a5}
\end{equation}
with the following (anti)commutation properties:
\begin{equation}
\{{\cal Q}_\alpha,{\cal Q}_\beta\}=
4i(P^2+m^2)(J\gamma)_{\alpha\beta}
-8i(\Delta_+\Pi_+
+\Delta_-\Pi_-)
(P\gamma)_{\alpha\beta},
\label{a6}
\end{equation}
\begin{equation}
[{\cal Q}_\alpha,{\cal Q}_\beta]=-
({\cal {\cal Q}}^\rho {\cal {\cal Q}}_\rho)
\epsilon_{\alpha\beta},
\quad
{\cal Q}^\rho {\cal Q}_\rho=i(P^2+m^2)(1+\nu R)
+8i\epsilon m
(\Delta_+\Pi_+
-\Delta_-\Pi_-),
\label{a7}
\end{equation}
where
$
\Delta_\pm=PJ-\epsilon m\frac{1}{4}(\nu\pm 1).
$
The list of relations has been used for calculating eqs.
(\ref{a6}), (\ref{a7}) is given in Appendix.  One introduces
the field $\Psi=\Psi^n(x)$, which carries the corresponding
irreducible representation of RDHA and is transformed
appropriately under the action of the Lorentz transformations
specified by parameters $\omega_\mu$,
$\Psi(x)\rightarrow \Psi'(x')=\exp(iM_\mu\omega^\mu)\Psi(x)$.
Now, let us postulate the spinor set of equations 
\begin{equation}
{\cal Q}_\alpha\Psi=0.
\label{a8}
\end{equation}
If the field $\Psi(x)$ satisfy equations (\ref{a8}),
in the Lorentz-transformed system it will satisfy the
equations ${\cal Q}'_\alpha\Psi'(x')=0$, 
${\cal Q}'_\alpha=\exp(iM^\mu\omega_\mu){\cal Q}_\alpha
\exp(-iM^\mu\omega_\mu)$. 
Therefore, the theory will have a covariant content.  
Due to eqs. (\ref{a6}), (\ref{a7}), we find that 
the fields $\Psi_\pm$,
$\Psi=\Psi_++\Psi_-$, $R\Psi_\pm=\pm\Psi_\pm$, 
satisfying eqs. (\ref{a8}),
will satisfy also quadratic Klein-Gordon and linear spin
equation of the  form (\ref{a1})
with parameter $s_+=\epsilon\frac{1}{4}(\nu+1)$
for the case of field $\Psi_+$ and 
$s_-=\epsilon\frac{1}{4}(\nu-1)$
for the field $\Psi_-$.
Then we can go over to the momentum representation and choose
the rest frame system $P^\mu=(\epsilon^0 m,0,0)$, $|\epsilon^0|=1$.  
As a result, we
find that the second equation from (\ref{a1})
has only trivial solution in the `$-$' subspace, i.e. 
$\Psi_-=0$.  On the other hand, one finds that the second
equation has nontrivial solutions in the `$+$' subspace with
$\epsilon^0$ taking values $+1$ and $-1$ 
in the case of finite-dimensional representations and
taking only the value $+1$ when $\nu>-1$. 
The corresponding  solution $\Psi_+$ describes irreducible
representations of the Poincar\'e group with $P^2=-m^2$ and spin
$s_+=\epsilon\frac{1}{4}(1+\nu)$.
The presence of the discrete parameter $\epsilon$
in operator (\ref{a5}) allows us to describe 
the states with spin of any sign.

Let us discuss the obtained result in more detail.
As we have seen, equations (\ref{a1}) appear as a consequence of
equations (\ref{a8}).  For the choice of finite dimensional
representations of deformed Heisenberg algebra  corresponding to
the values of deformation parameter $\nu=-3$ and $\nu=-5$, the
second equation from (\ref{a1}) taken for 
corresponding nontrivial field $\Psi_+$ 
is nothing else as the Dirac or
Deser-Jackiw-Templeton-Schonfeld (DJTS)
equation for the topologically massive vector gauge field
\cite{djts}.  Therefore, we  conclude that the constructed set of
equations (\ref{a8}) for the choice of the parameter $\nu=-3$ and
$\nu=-5$ is equivalent either to the Dirac equation or to the
DJTS equation.  For other choices of finite-dimensional
representations of RDHA $(\nu=-(2r+1)$, $r=3,4,\ldots$), the
basic set of equations (\ref{a8}) describes irreducible
representations of the $d=2+1$ Poincar\'e group with integer,
$s_+=-\epsilon k$ ( for $r=2k$), and half integer,
$s_+=-\epsilon (k+1/2)$ (for $r=2k+1$), spin values.  
For infinite-dimensional representations of RDHA,
our spinor set (\ref{a8}) is some (2+1)-dimensional
analog of the Dirac positive-energy set of equations \cite{dir}
related to the infinite-component Majorana positive-energy
equation \cite{maj,gr2}.
Indeed, in the momentum representation we get the 
positive energy solution
$\Psi_+$ carrying arbitrary nonzero spin 
$s_+=\epsilon\alpha_+$, $\alpha_+=\frac{1}{4}(1+\nu),$
\begin{equation}
\Psi^n_{+\bf k}(P)=
(-1)^n C^{1/2}_n
\left(\frac{P_1-iP_2}{P^0+m}\right)^{n}
\Psi^0_{+\bf k}(P),\quad
\Psi^0_{+\bf k}(P)=
\delta(P^0-\sqrt{{\bf P}^2+m^2})\delta ({\bf P}-
{\bf k})f({\bf k}),
\label{a9}
\end{equation}
where $C_n=\Gamma(2\alpha_++n)/\Gamma(2\alpha_+)\Gamma(n+1)$
and $f({\bf k})$ is an arbitrary function.

Therefore, the constructed minimal spinor set of linear
differential equations (\ref{a8}) has a universal character
describing spin-$j$ fields and anyons.

On the even subspace `$+$' spanned by the states 
of the form $|k\rangle_+\propto (a^+)^{2k}|0\rangle$, 
the following relation takes place:
$
-\frac{i}{4}{\cal L}^\alpha(\gamma_\mu)_{\alpha\beta}{\cal
Q}^{\beta}=
\frac{1}{4}(1+\nu)P_\mu+\epsilon mJ_\mu-i\epsilon_{\mu\nu\lambda}
P^\nu J^\lambda\equiv V_\mu.
$
Operator $V_\mu$ is exactly the operator 
generating the universal vector set of linear differential equations
$V_\mu\Psi=0$ constructed in ref. \cite{gr3}. However, 
due to the operator identity
$T^\mu V_\mu\equiv0$,
$T_\mu=(\frac{1}{4}(\alpha_+-1)^2\eta_{\mu\nu}-
i(\alpha_+-1)\epsilon_{\mu\nu\lambda}J^\lambda +J_\nu
J_\mu)P^\nu$, 
the vector set, unlike the present spinor case, is the  set of
dependent equations.  This fact made the problem of construction
of the corresponding field action to be complicated \cite{gr3}.

The spinor operator ${\cal Q}_\alpha$, unlike the operator
${\cal D}_\alpha=\epsilon m{\cal L}_\alpha -{\cal P}_\alpha\neq
{\cal D}^\dagger_\alpha$ used
earlier for constructing the minimal set of anyonic equations
\cite{gr4,gr6}, is self-conjugate, ${\cal Q}^\dagger_\alpha={\cal
Q}_\alpha$.  As we shall see, this property of our basic spinor
operator will be important for formulating the problem of
construction of the classical spin model corresponding to the
universal
field system (\ref{a8}).

\section{(2+1)-dimensional SUSY}
Any irreducible representation of the constructed $osp(1|2)$
superalgebra contains the direct sum of two irreducible 
representations of $so(2,1)$ subalgebra, which are  
specified by the parameters $\alpha_+$ and $\alpha_-$ or
$j_+$ and $j_-$ related as $\alpha_- -\alpha_+=j_+-j_-=1/2$.
Therefore, it seems rather natural to try to modify the set of
linear differential equations in such a way that it would have
nontrivial solutions not only in `$+$' subspace, but also in
`$-$' subspace. If these states will have equal mass but their
spin will be shifted for $\Delta s=1/2 ({\rm mod}\, n)$, we shall
have the states forming a supermultiplet.  As we shall see, it
is indeed possible to find such a modification of spinor set of
linear differential equations, but a program turns out to
be realizable only for some special cases of finite-dimensional
representations of RDHA.

In order to realize such a modification,
let us consider the spinor set of equations
\begin{equation}
{\cal D}_\alpha\Psi=0
\label{a10}
\end{equation}
with spinor linear differential operator
of the most general form,
$
{\cal D}_\alpha={\cal D}^+_\alpha \Pi_++{\cal D}^-_\alpha\Pi_-,
$
where
$
{\cal D}^\pm_\alpha={\cal P}_\alpha A^\pm +{\cal J}_\alpha B^\pm
-m{\cal L}_\alpha C^\pm,
$
and $A^\pm$, $B^\pm$ and $C^\pm$ are $c$-number constants.
Contracting  equations (\ref{a10}) with ${\cal L}^\alpha$
(see Appendix),
we get
\begin{equation}
\left(PJ(A^\pm+B^\pm)-m\frac{1}{4}(1\pm \nu)C^\pm\right)
\Psi_\pm=0.
\label{a11}
\end{equation}
In order this would give spin equations 
$(PJ-ms_\pm)\Psi_\pm=0$,
one has to have
$
A^\pm+B^\pm\neq 0
$
and
$
\frac{1}{4}(1\pm\nu)C^\pm=s_\pm(A^\pm+B^\pm).
$
Then, the contraction of eq. (\ref{a10}) with ${\cal P}^\alpha$
gives 
\begin{equation}
F^\pm\left(P^2+m^2\frac{16s_\pm^2}{(1\pm\nu)^2}\right)\Psi_\pm=0,
\quad
F^\pm=A^\pm+\frac{1}{4}(3\mp\nu)B^\pm.
\label{a12}
\end{equation}
At last, the contraction with ${\cal J}^\alpha$ gives 
conditions (\ref{a12}) multiplied by $(1\pm\nu)^2$.
If $F^+\neq0$ and $A^++B^+\neq0$, two conditions (\ref{a11})
and (\ref{a12}) are consistent on the `$+$' subspace
for $|s_+|=\frac{1}{4}(1+\nu)$,
and as a consequence, for $\Psi_+$
we have  the Klein-Gordon and spin 
equations of the form (\ref{a1}).
On the other hand if $F^-\neq0$,
corresponding conditions for `$-$' subspace have only trivial
solution $\Psi_-=0$. Let 
$F^-=0$, $B^-\neq0$ and $C^-=s_-B_-$.
Then we have ${\cal D}^+_\alpha={\cal P}_\alpha A^++{\cal J}_\alpha
B^++ \epsilon m {\cal L}_\alpha (A^++B^+)$, $\epsilon=\pm1$,
and ${\cal D}^-_\alpha=(-\frac{1}{4}(3+\nu){\cal P}_\alpha
+{\cal J}_\alpha -ms_-{\cal L}_\alpha)B^-$.
As a result,
in this case eqs. (\ref{a10})
lead to  equations (\ref{a1}) for 
the `$+$' subspace,  whereas for `$-$' subspace 
we have only one equation,
$
(PJ-ms_-)\Psi_-=0.
$
This equation itself generally does not
fix the value of the Casimir operator $P^2$
neither for the case $\nu>-1$ nor for the case 
$\nu=-(2r+1)$.
However, there are two exceptional cases
given by $\nu=-5$ and $\nu=-7$.
In these two cases 
the mass shell condition $(P^2+m^2)\Psi_-=0$
appears on the `$-$' subspace as a consequence of
the spin equation being the Dirac or DJTS equation.  

Let us require that the spinor operator ${\cal D}_\alpha$
would be self-conjugate, i.e.
${\cal D}^\dagger_\alpha={\cal D}_\alpha$.
This requirement leads to the conditions $(B^-)^*=B^+$, 
$A^+=-\frac{1}{4}(1-\nu)B^+$, and to the relation
$
s_-=s_+(3+\nu)/(1+\nu).
$
In the case $\nu=-5$, we get $s_+=\epsilon$,
$s_-=\frac{1}{2}s_+$, i.e.
we have a massive supermultiplet of states with 
the spin content $(s_+=\epsilon,$ $s_-=\frac{1}{2}\epsilon)$.
For $\nu=-7$, we have a supermultiplet
with the spin content $(s_+=\frac{3}{2}\epsilon,$
$s_-=\epsilon$). 
Putting $B^-=-1$,
for the cases $\nu=-5$ ($r=2$) and $\nu=-7$ ($r=3$),
we have equation (\ref{a10}) with 
\begin{equation}
{\cal D}_\alpha=\frac{r-1}{2}\epsilon m{\cal L}_\alpha-{\cal
J}_\alpha
+\frac{1}{2}(1-rR){\cal P}_\alpha,\quad
r=2,3.
\label{a13}
\end{equation}
It is interesting to note that in both cases, $r=2,3,$
the spinor supercharge operator
${\cal Q}_\alpha$, $\{{\cal Q}_\alpha, {\cal D}_\beta\}\approx 0$,
which transforms the corresponding supermultiplet components, 
coincides with operator (\ref{a5}).
The weak equality means
that the left hand side turns into zero on the 
physical subspace given by eqs. (\ref{a10}),
and we have the following
typical superalgebraic relations: 
$
\{{\cal Q}_\alpha,{\cal Q}_\beta\}
\approx-4i\epsilon m
(r+2)(P\gamma)_{\alpha\beta},
$
$
[P_\mu,{\cal Q}_\alpha]=0.
$

If we not require that
${\cal D}_\alpha^\dagger={\cal D}_\alpha$,
we shall have a possibility to get supermultiplets with exotic 
spin content
$(s_+=\epsilon,$ $s_-=-\frac{1}{2}\epsilon)$ 
and with corresponding spin shift $|\Delta s|=|s_+-s_-|=3/2$ 
for $\nu=-5$,
and $(s_+=\frac{3}{2}\epsilon,$ $s_-=-\epsilon$) 
with $|\Delta s|=5/2$ for $\nu=-7$.
Choosing $B^+=B^-=-1$, $A^+=1/2$ for $\nu=-5$,
and $B^+=B^-=-1/2$, $A^+=0$ for $\nu=-7$,
we shall obtain 
\begin{equation}
{\cal D}_\alpha=-\frac{1}{2}\epsilon m{\cal L}_\alpha-{\cal J}_\alpha
-\frac{1}{2}R{\cal P}_\alpha,\quad \nu=-5;
\quad
{\cal D}_\alpha=-\frac{1}{2}
\epsilon m{\cal L}_\alpha-\frac{1}{2}{\cal J}_\alpha
-\frac{1}{4}(1+R){\cal P}_\alpha,\quad \nu=-7.
\label{a13*}
\end{equation}
Both these operators have the property
${\cal D}^\dagger_\alpha={\cal D}_\alpha+{\cal P}_\alpha$.

\section{Field action}
Since universal equations (\ref{a8}) 
are two equations for one field
$\Psi(x)$, the corresponding field action 
giving rise to them has to contain some auxiliary fields.
Let us consider the action
\begin{equation}
S=\int Ld^3x
\label{a14}
\end{equation}
with Lagrangian
$
L={\bar \chi}{}^{\alpha}{\cal Q}_\alpha {\cal Q}_\beta\chi^\beta+
{\bar \chi}{}^\alpha {\cal Q}_\alpha\Psi+
{\bar \Psi}{\cal Q}_\alpha \chi^\alpha +c{\bar \Psi}\Psi,
$
containing auxiliary field $\chi^\alpha$
with dimensionality $[\chi^\alpha]=m^{1/2}$,
whereas $[\Psi]=m^{3/2}$, and here $c=c^*$ is a real dimensionless
constant.
The conjugate fields are
${\bar \Psi}=\Psi^\dagger$,
${\bar \chi}{}^{\alpha}=\chi^{\dagger}{}^{\alpha}$
for $\nu>-1$, whereas
${\bar \Psi}=\Psi^\dagger\hat{\eta}$,
${\bar \chi}{}^{\alpha}=\chi^{\dagger}{}^{\alpha}
\hat{\eta}$ for $\nu=-(2r+1)$ that
guarantees the reality of the Lagrangian.
Action (\ref{a14}) leads to the equations
$
{\cal Q}_\alpha {\cal Q}_\beta\chi^{\beta}+{\cal Q}_\alpha\Psi=0,
$
$
{\cal Q}_\alpha\chi^\alpha+c\Psi=0,
$
and to the corresponding equations for the conjugate fields
${\bar \Psi}$ and ${\bar \chi}{}^\alpha$.
Substituting the second equation into the first,
we arrive at the equivalent system
of equations
$
{\cal Q}_\alpha(1-c)\Psi=0,
$
$
{\cal Q}_\alpha\chi^\alpha+c\Psi=0.
$
Therefore, if $c\ne 1$, we have the necessary spinor
set of equations (\ref{a8}) for the basic field $\Psi$.
Now, we have to arrange the construction in such
a way that the field $\chi^\alpha$
would be a pure auxiliary field having no independent dynamics.
To do this, we decompose the field
$\chi^\alpha$ into ``longitudinal" and ``transverse" parts,
$\chi^\alpha={\cal Q}^\alpha\chi_l+\chi^\alpha_\bot$,
where ${\cal Q}_\alpha\chi^\alpha=\chi_l$, and 
${\cal Q}_\alpha\chi^\alpha_{\bot}=0$.
Due to the equations of motion, 
the ``longitudinal" part of $\chi^\alpha$ 
coincides up to a $c$-number factor with $\Psi$.
If we choose $c=0$, the ``longitudinal" part
of $\chi^\alpha$ will not contribute at all to the theory.
On the other hand, the ``transverse" part
of $\chi^\alpha$ is a pure gauge degree of freedom
of the theory
due to the invariance of the action and of the equations of motion
with respect to the transformations
$\chi^\alpha\rightarrow \chi^\alpha+\Pi^{\alpha\beta}\Lambda_\beta$,
where 
$
\Pi_{\alpha\beta}=({\cal Q}^\sigma {\cal
Q}_\sigma)\epsilon_{\alpha\beta}
-{\cal Q}_\alpha {\cal Q}_\beta,
$
$
{\cal Q}^\alpha\Pi_{\alpha\beta}=0.
$
Therefore, at $c=0$, $\chi^\alpha(x)$ plays a role of an
auxiliary spinor field.  Finally, the Lagrangian can be represented
in
the form
\begin{equation}
L=({\bar \Psi}+{\bar \chi}{}^\alpha {\cal Q}_\alpha)
(\Psi+{\cal Q}_\beta\chi^\beta)-{\bar \Psi}\Psi.
\label{a16}
\end{equation}
Note that the equation ${\cal Q}_\alpha\chi^\alpha=0$ taking
place for the choice $c=0$ in the initial Lagrangian, together
with the basic spinor set of linear differential equations
(\ref{a8}) may be obtained from the action with the linear form of
Lagrangian,
\begin{equation}
L'={\bar \chi}{}^\alpha {\cal Q}_\alpha\Psi+
{\bar \Psi}{\cal Q}_\alpha\chi^\alpha.
\label{a17}
\end{equation}

The change of ${\cal Q}_\alpha$ for ${\cal D}_\alpha$ in
Lagrangians (\ref{a16}) and (\ref{a17}) will give the action
functionals for the special SUSY cases (\ref{a13}) 
corresponding to $\nu=-5$ and $\nu=-7$.
For SUSY systems given by eqs. (\ref{a10}),
(\ref{a13*}), the corresponding quadratic and linear
Lagrangians are
$
L=({\bar \Psi}+{\bar \chi}{}^\alpha {\cal D}_\alpha)
(\Psi+{\cal D}^\dagger_\beta\chi^\beta)-{\bar \Psi}\Psi
$
and 
$
L'={\bar \chi}{}^\alpha {\cal D}_\alpha\Psi+
{\bar \Psi}{\cal D}^\dagger_\alpha\chi^\alpha.
$

\section{Concluding remarks}
In the approach with Chern-Simons gauge field, a gauge-invariant
anyonic composite field is a nonlocal field of  the initial
statistically charged matter field and it is given on a
half-infinite non-observable space-like string \cite{cs1,cs2}.  
In the present approach, we describe fractons by infinite-component
field with index $n$ taking half-infinite set of values,
$n=0,1,\ldots$.  At the same time, in accordance with eq.
(\ref{a9}) this infinite-component field carries only one
independent field degree of freedom.  Therefore, there is some
analogy of the group-theoretical approach with the Chern-Simons
approach.  The spin of fractons can be varied continuously by
changing the deformation parameter $\nu$.  In particular, it can
take integer or half-integer values. However, in such cases the
description of integer and half-integer spin fields is
essentially different from the description of spin-$j$ fields of
the same spin value: spin-$j$ fields are described by
non-unitary finite-dimensional $so(2,1)$ representations,
whereas fracton case corresponds to employing unitary
infinite-dimensional representations.  Perhaps, the above
mentioned analogy means that the description of integer and
half-integer relativistic spin fields given by the Chern-Simons
approach under the appropriate choice of the statistical
parameter is  not reducible to the ordinary description of
spin-$j$ fields (see, however, ref. \cite{chit}).

The quantization of the theory presented here deserves a detailed 
Hamiltonian analysis. As we noted,
in the case of the choice of `anyonic representation'
of the $R$-deformed Heisenberg algebra 
($\nu>-1$), infinite-component field carries effectively only one 
independent physical degree of freedom.
Therefore, corresponding Hamiltonian
theory has to contain infinite set of constraints removing
all the field degrees of freedom except one. 
Such infinite set of constraints should
appropriately be taken into account.
The subsequent quantization of the theory 
given by Lagrangians (\ref{a16}) and (\ref{a17}) has to 
answer the question on the spin-statistics relation for fractons.

Our final remark concerns the construction of classical
mechanical model which would 
correspond to the field system defined by eqs. (\ref{a5}),
(\ref{a8}).
Due to the property ${\cal Q}^\dagger_\alpha={\cal Q}_\alpha$,
one can interpret equations (\ref{a8}) as quantum analogs
of some classical constraints $q_\alpha\approx 0$
which should be supplemented by the constraints
$p^2+m^2\approx 0$ and $pj-ms\approx 0$ being the 
classical analogs of the Klein-Gordon and spin equations.
In this case reflection operator should be understood 
as the operator realized 
via creation-annihilation operators 
$a^\pm$ as $R=\cos \pi N$,
$N=\frac{1}{2}\{a^+,a^-\}-\frac{1}{2}(1+\nu)$.
Therefore, in this case spinor operators (\ref{a5})
can be understood as essentially nonlinear operators 
in terms of operators $a^\pm$. 
Then, the problem will be reduced to the problem of
constructing a classical analog of
algebra (\ref{a3}) with $R=\cos\pi N$.
It seems that the classical analog of this algebra can be constructed
using the ideas of ref. \cite{sha}
where the classical analog of the $q$-deformed
oscillator was realized as a constrained system given on a 
K\"ahler manifold. 
The construction of such universal classical mechanical
model for $d=2+1$ spin systems could help in resolving 
some open problems taking place in pseudoclassical 
approach to relativistic spin-$j$ particles \cite{prob}.
\vskip0.3cm

{\bf Acknowledgements}

The author thanks the Department of Physics of the Federal
University of Juiz de Fora, where the part of this work
was done, for hospitality. 

\appendix

\section{Appendix}
The following contraction relations take place for the operators 
${\cal L}_\alpha$, ${\cal P}_\alpha$ and ${\cal J}_\alpha$:
\begin{eqnarray}
&{\cal L}^\alpha {\cal L}_\alpha=-i(1+\nu R),\quad
{\cal L}^\alpha{\cal P}_\alpha=
-{\cal P}^\alpha {\cal L}_\alpha=
{\cal L}^\alpha {\cal J}_\alpha=
=-4i(PJ),\quad
{\cal J}^\alpha {\cal L}_\alpha=0,
&
\nonumber\\
&
{\cal P}^\alpha {\cal P}_\alpha=-i(1+\nu R)P^2,
\quad
{\cal J}^\alpha {\cal P}_\alpha=4i\left(P^2 J^2-(PJ)^2\right)
-i(1+\nu R)P^2,
&
\nonumber\\
&
{\cal P}^\alpha {\cal J}_\alpha=4i\left((PJ)^2-P^2
J^2\right),
\quad
{\cal J}^\alpha {\cal J}_\alpha=i(1+\nu R)\left(
(PJ)^2-P^2 J^2 \right)-4i(PJ)^2.
&
\nonumber
\end{eqnarray}
The complete set of (anti)commutation relations
for these operators is the following:
\[
[{\cal S}_\alpha,{\cal S}_\beta]=-({\cal S}^\rho
{\cal S}_\rho)\epsilon_{\alpha\beta},\quad
{\cal S}_\alpha={\cal L}_\alpha,{\cal P}_\alpha,
{\cal J}_\alpha,
\]
\[
\{{\cal L}_\alpha,{\cal L}_\beta\}=4i(J\gamma)_{\alpha\beta},
\quad
\{{\cal P}_\alpha,{\cal P}_\beta\}=
8i(PJ)(P\gamma)_{\alpha\beta}
-4iP^2(J\gamma)_{\alpha\beta},
\]
\[
[{\cal L}_\alpha,{\cal P}_\beta]
=-i(1+\nu R)(P\gamma)_{\alpha\beta},
\quad
\{{\cal L}_\alpha,{\cal P}_\beta\}
=4i(PJ)\epsilon_{\alpha\beta}
-4\Gamma_{\alpha\beta},\quad
\Gamma_{\alpha\beta}\equiv
\epsilon_{\mu\nu\lambda}P^\mu J^\nu \gamma^\lambda_{\alpha\beta},
\]
\[
[{\cal L}_\alpha,{\cal J}_\beta]
=-\frac{i}{2}(1+\nu R)
(P\gamma)_{\alpha\beta}+(1+\nu R)
\Gamma_{\alpha\beta}
+2i(PJ)\epsilon_{\alpha\beta},
\]
\[
\{{\cal L}_\alpha,{\cal J}_\beta\}=
-4\Gamma_{\alpha\beta}
+2i(PJ)\epsilon_{\alpha\beta}+4i(J\gamma)_{\alpha\beta}(PJ)
+\frac{i}{4}(\nu^2-1)(P\gamma)_{\alpha\beta},
\]
\[
[{\cal P}_\alpha,{\cal J}_\beta]=\frac{i}{2}
(1+\nu R)P^2\epsilon_{\alpha\beta}
+i(1+\nu R)P^2(J\gamma)_{\alpha\beta}+i
(1-\nu R)(PJ)(P\gamma)_{\alpha\beta},
\]
\[
\{{\cal P}_\alpha,{\cal J}_\beta\}=-4iP^2(J\gamma)_{\alpha\beta}
+6i(PJ)(P\gamma)_{\alpha\beta}
-4\Gamma_{\alpha\beta}(PJ)
+\frac{i}{4}\left((1-\nu^2)P^2
-16(PJ)^2\right)\epsilon_{\alpha\beta},
\]
\[
\{{\cal J}_\alpha,{\cal J}_\beta\}=
4i (J\gamma)_{\alpha\beta}\left((PJ)^2-P^2+
\frac{1}{16}(1+\nu R)^2 P^2\right)+
4i(PJ)(P\gamma)_{\alpha\beta}
-8\Gamma_{\alpha\beta}(PJ).
\]


\begin{thebibliography}{**}

\bibitem{a1}
J.M. Leinaas and J. Myrheim,
{\it Nuovo Cimento} {\bf B37} (1977) 1;\\
G.A. Goldin, R. Menikoff and D.H. Sharp,
{\it J. Math. Phys.} {\bf 22} (1981) 1664;\\
F. Wilczek, {\it Phys. Rev. Lett.} {\bf 48} (1982) 1144;
{\bf 49} (1982) 957

\bibitem{a2}
B. Binegar, {\it J. Math. Phys.} {\bf 23} (1982) 1511

\bibitem{a3}
R. Mackenzie and F. Wilczek, {\it Int. J. Mod. Phys.} {\bf A3}, 
2827 (1988);\\
J. Fr${\rm \ddot{o}}$hlich and P. Marchetti,
{\it Nucl. Phys.} {\bf B35} (1991) 533;\\
S. Forte, {\it Rev. Mod. Phys.} {\bf 64} (1992) 193;\\
K. Lechner and R. Iengo, {\it Phys. Rep.} {\bf 213} (1992) 179
\bibitem{cs1}
G.W. Semenoff, {\it Phys. Rev. Lett.} {\bf 61} (1988) 517

\bibitem{cs2}
R. Banerjee, A. Chatterjee and V.V. Sreedhar,
{\it Ann. Phys.} {\bf 222} (1993) 254

\bibitem{gr1}
M.S. Plyushchay, {\it Phys. Lett.} {\bf B248} (1990) 107

\bibitem{jn}
R. Jackiw and V.P. Nair, {\it Phys. Rev.} {\bf D43} (1991) 1933

\bibitem{gr2}
M.S. Plyushchay, {\it Phys. Lett.} {\bf B262} (1991) 71;
{\it Nucl. Phys.} {\bf B362}  (1991) 54

\bibitem{sv}
D.P. Sorokin and D.V. Volkov, {\it Nucl. Phys.}  {\bf B409} (1993)
547
\bibitem{gr3}
J.L. Cort\'es and M.S. Plyushchay,
{\it J. Math. Phys.} {\bf 35} (1994) 6049

\bibitem{gr4}
M.S. Plyushchay, {\it Phys. Lett.} {\bf B320} (1994) 91

\bibitem{gr5}
J.L. Cort\'es and M.S. Plyushchay,
{\it Int. J. Mod. Phys.} {\bf A11} (1996) 3331

\bibitem{gr6}
M. S. Plyushchay,
{\it Ann. Phys.} {\bf 245} (1996) 339

\bibitem{chou}
C. Chou, V.P. Nair and A.P. Polychronakos,
{\it Phys. Lett.} {\bf B304} (1993) 105;\\
Ch. Chou, {\it Phys. Lett.} {\bf B323} (1994) 147

\bibitem{pseud}
L. Brink, S. Deser, B. Zumino, P. Di Vecchia and P. Howe,
{\it Phys. Lett.} {\bf 64} (1976) 435;\\
R. Casalbuoni, {\it Nuovo Cimento} {\bf A33} (1976) 369;\\
F.A. Berezin and M.S. Marinov, 
{\it Ann. Phys.} {\bf 104} (1977) 336

\bibitem{para}
G.P. Korchemsky, {\it Phys. Lett.} {\bf 267B} (1991) 497

\bibitem{ok}
Y. Ohnuki and S. Kamefuchi,
{\it Quantum Field Theory and Parastatistics}
(University Press of Tokyo, 1982);\\
A.J. Macfarlane, 
{\it Generalized Oscillator Systems and Their 
Parabosonic Interpretation}, in:
Proc. Inter. Workshop on Symmetry Methods in Physics,
eds. A.N. Sissakian, G.S. Pogosyan and 
S.I. Vinitsky (JINR, Dubna, 1994) p. 319

\bibitem{uni}
M.S. Plyushchay,
{\it Mod. Phys. Lett.} {\bf A11} (1996) 2953;
{\it Nucl. Phys.} {\bf B491} (1997) 619 
[hep-th/9701091].

\bibitem{dir}
P.A.M. Dirac, {\it Proc. Roy. Soc. London Ser. A}
{\bf 322} (1971) 435, {\bf 328} (1972) 1572 

\bibitem{fiks}
A.T. Filippov, A.P. Isaev, and A.P. Kurdikov,
{\it Mod. Phys. Lett.} {\bf A7} (1992) 2129;
{\it Teor. Mat. Fiz.} {\bf 94} (1993) 213;\\
V. Spiridonov, {\it Parasupersymmetry in Quantum Systems,}
in Proc. of the XXth Intern. Conf. on Differential Geometric
Methods in Theor. Phys. (June 3-7, 1991, New York, USA),
Eds. S.Catto and A.Rocha, World Sci. 1992, vol.1, p. 622;\\
A.P. Isaev, {\it Paragrassmann Integral, Discrete Systems and
Quantum Groups}, Preprint q-alg/9609030

\bibitem{djts}
R. Jackiw and S. Templeton, {\it Phys. Rev.} {\bf D23} (1981) 2291\\
J. Schonfeld, {\it Nucl. Phys.} {\bf B185} (1981) 157;\\
S. Deser, R. Jackiw, and S. Templeton,
{\it Phys. Rev. Lett.} {\bf 48} (1982) 975
\bibitem{maj}
E. Majorana, {\it Nuovo Cim.} {\bf 9} (1932) 335

\bibitem{chit}
W. Chen and Ch. Itoi, {\it Nucl. Phys.} {\bf B435} (1995) 690

\bibitem{sha}
S.V. Shabanov, {\it Mod. Phys. Lett.} {\bf A10} (1995) 941

\bibitem{prob}
J.L. Cort\'es, M. S. Plyushchay and L. Vel\'azquez,
{\it Phys. Lett.} {\bf B306} (1993)  34;\\
D.M. Gitman, A.E. Gon\c calves and I.V. Tyutin,
{\it Phys. Rev.} {\bf D50} (1994) 5439;\\ 
J.L. Cort\'es and M.S. Plyushchay, 
{\it Comment on "New Pseudoclassical Model for Weyl Particles"},
Preprint hep-th/9602106;\\
D.M. Gitman, {\it Path Integrals and Pseudoclassical Description 
for Spinning Particles in Arbitrary Dimensions},
Preprint hep-th/9608180

\end{thebibliography}
\end{document}